\def\apj{ApJ}

\def\apjl{{ApJL}}

\def\aap{{AAP}}

\def\taxp{\hbox{XTE~J1810$-$197}}

\newcommand\simgt{\lower.5ex\hbox{$\; \buildrel > \over \sim \;$}}

\newcommand\xte{{RXTE}}

\newcommand\rosat{{ROSAT}}
\newcommand\chandra{{Chandra}}

\newcommand\xmm{{XMM-Newton}}

\documentclass[twocolumn,runningheads,natbib]{svjour2}
\smartqed  
\usepackage{graphicx,psfig,rotating}
\journalname{Astrophysics and Space Science}
\begin{document}

\title{
The Anatomy of a Magnetar: XMM Monitoring of the Transient Anomalous X-ray Pulsar XTE~J1810--197
\thanks{\xmm\ is an ESA science mission with instruments and contributions directly funded by ESA
Member States and NASA. This research is supported by \xmm\ grant
NNG05GJ61G and NASA ADP grant ADP04-0059-0024.}
}


\author{E. V. Gotthelf \and J.~P.~Halpern
}


\institute{E.~V.~Gotthelf \at
Columbia Astrophysics Laboratory, Columbia
University, 550 West 120$^{th}$ Street, New York, NY 10027-6601\\
              \email{eric@astro.columbia.edu}           
}

\date{Received: date / Accepted: date}

\maketitle

\begin{abstract}

We present the latest results from a multi-epoch timing and spectral
study of the Transient Anomalous X-ray Pulsar XTE~J1810--197. We have
acquired seven observations of this pulsar with the Newton X-ray
Multi-mirror Mission (\xmm) over the course of two and a half years,
to follow the spectral evolution as the source fades from
outburst. The spectrum is arguably best characterized by a
two-temperature blackbody whose luminosities are decreasing
exponentially with $\tau_1 = 870$~d and $\tau_2 = 280$~d,
respectively. The temperatures of these components are currently
cooling at a rate of 22\% per year from a nearly constant value
recorded at earlier epochs of $kT_1 = 0.25$~keV and $kT_2 = 0.67$~keV,
respectively.  The new data show that the temperature $T_1$ and
luminosity of that component have nearly returned to their historic
quiescent levels and that its pulsed fraction, which has steadily
decreased with time, is now consistent with the previous lack of
detected pulsations in quiescence. We also summarize the detections of
radio emission from XTE~J1810--197, the first confirmed for any
AXP. We consider possible models for the emission geometry and
mechanisms of XTE~J1810--197.
\keywords{pulsars: general --- stars: individual (XTE J1810--197) --- 
stars: neutron --- X-rays: stars}
\PACS{95.85.-e \and\ 95.85.Nv \and\ 97.60.Jd \and\ 97.60.Gb \and\ 95.75}
\end{abstract}

\vskip 0.5cm
\section{Introduction}

Neutron star (NS) astronomy has been recently invigorated by the
identification of a new class of magnetically dominated emitters.
Known as anomalous X-ray pulsars (AXPs) and soft gamma-ray repeaters
(SGRs), these objects are apparently young, isolated neutron stars
(NSs), whose properties differ markedly from those of the Crab pulsar,
previously considered prototypical of the young NSs (for a review see
\citealt{mer02}). All AXPs and at least 3 of the 4 SGRs are 
identified as relatively slow ($5-12$~s) pulsars. Many are located at
the centers of recognized, young supernova remnants (SNRs), directly
associating them with their supernova explosions. These objects emit
predominantly at X-ray energies and are distinguished by their
characteristic spectral signature. This radiation cannot be accounted
for by rotational energy losses alone, as for the radio pulsars, but
is most likely powered by the decay of an enormous magnetic field
characterized by a dipole with $B_p \simgt 4.4\times 10^{13}$~G,
at the pole.  Collectively, these isolated NSs are understood within
the context of the magnetar theory (\citealt{dun92,tho96}).

A unique pulsar has been discovered whose remarkable
properties offer great promise for deciphering the emission
mechanism(s) of magnetars. \taxp\ is a 5.54~s X-ray pulsar whose
measured and derived physical parameters are fully characteristic of
an AXP, but expresses behavior not previously associated with any such
object \citep[hereafter Paper~I]{got04}. \taxp\ is a transient AXP
(TAXP) -- it was discovered during a bright impulsive outburst
\citep{ibr04} that is still fading steadily. Even more surprising is
the discovery of highly variable radio emission, providing the first
confirmed example of radio flux from an AXP \citep{hal05b}, and the
subsequent detection of radio pulsations at the X-ray period
\citep{cam06}.

The outburst that resulted in the detection of the transient AXP occurred
sometime between 2002 November and 2003 January \citep{ibr04}. Since
then, over the course of a year, regular scans of the region with
\xte\ recorded an exponential flux decay with a time-constant of 
$269\pm25$~days from a maximum of $F (2-10\ {\rm keV}) \approx 8 
\times 10^{-11}$ erg~cm$^{-2}$~s$^{-1}$. In comparison, the previous 
average quiescent flux, with its softer spectrum, gives $F(0.5-10\
{\rm keV}) \approx 5.5 \times 10^{-13}$ erg~cm$^{-2}$~s$^{-1}$
(Paper~I).  This contrast is unprecedented for an AXP.  Further \xte\
observations yielded SGR-like episodes of bursts \citep{woo05},
similar to those seen from 1E~2259+586 \citep{kas03}. A search for an
optical/IR counterpart detected a fading IR source within the
\chandra\ error circle, similar to ones associated with other AXPs,
confirming its identification with \taxp\
\citep{rea04a,rea04b}.

In this paper we present the latest results from TAXP \taxp\ including
new \xmm\ observations extending to 2006 March. These observations
allow us to characterize the spectral evolution of a TAXP in outburst
as it returns to quiescence.  We show that the spectrum has now begun
a marked transition back to the nominal quiescent state, with a
distinct temperature and flux evolution. This long-term evolution
provides the unique possibility of decomposing its fading spectral
components in a manner unavailable for other magnetars
(\citealt[hereafter Paper~II]{hal05}). 
In the following, we used a
revised distant to the NS, $d_{3.3}$, quoted in units of $3.3$~kpc and
based on the radio pulsar DM measurement of
\cite{cam06}.

\section{Long-term spectral and temporal evolution}

Herein we present a total of seven \xmm\ observations of \taxp\ of
which four have been previously described in Papers~I, II, and
\citet[hereafter Paper~III]{got05}.  The three new data sets were 
obtained using similar observing modes as for previous observations
and reduced and analyzed in an identical manner. A log of these
observations is recorded in Table~1. A full report on the reduction
and analysis of these data sets will be presented in Gotthelf~\etal\
(in prep.).

\begin{figure}
\centering
\includegraphics[angle=270,width=0.45\textwidth,clip=true]{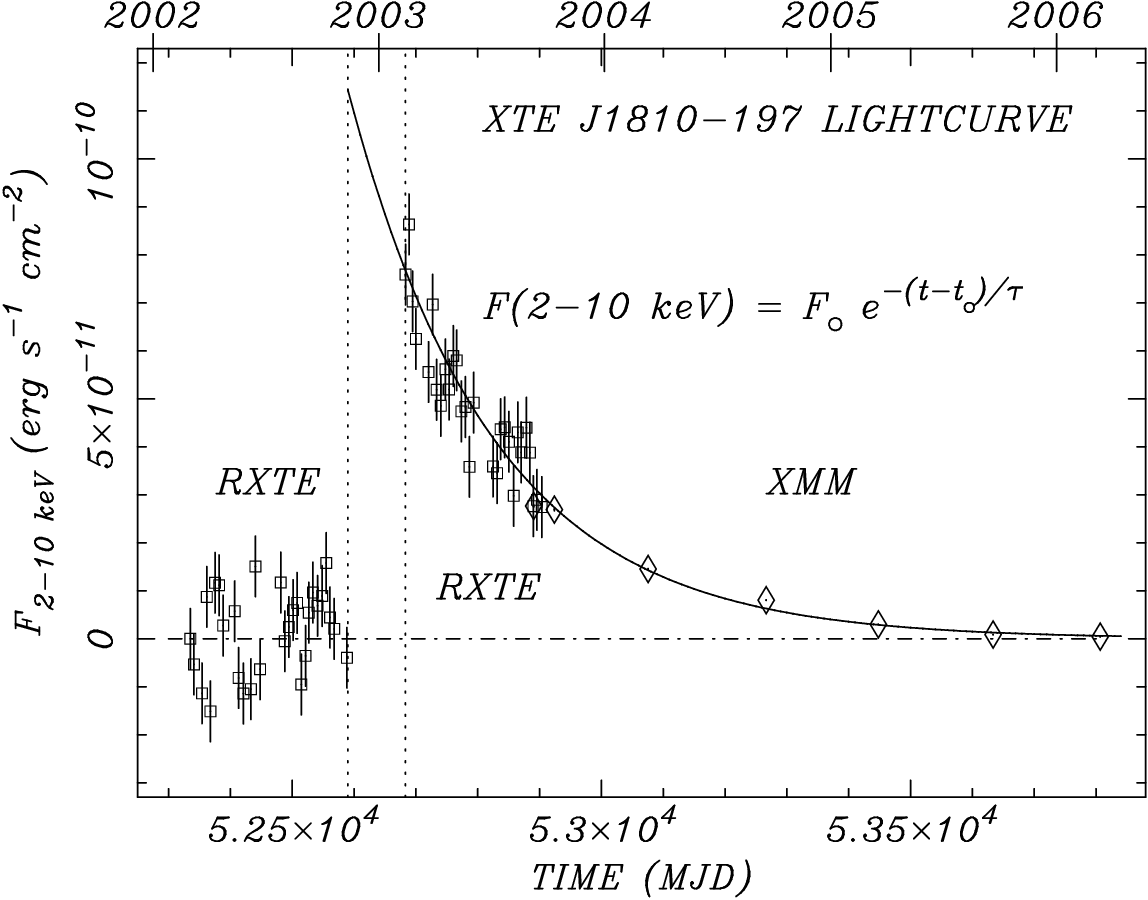}
\caption{Long-term light-curve of \taxp\ using  measurements obtained
with \xte\ ({\it squares}) and \xmm\ ({\it diamond}). The \xte\ data
are from \cite{ibr04}, renormalized to match the \xmm\ results.  The
\xmm\ data points are for the observations of Table~1, but fitted with
a power-law model in the $2-10$~keV energy band, for direct comparison
with the \xte\ results. The {\it solid line} is a best combined fit to
an exponential decay model (see text for parameters).  }
\label{fig:flux}
\end{figure}

Figure~\ref{fig:flux} presents an up-to-date light curve of TAXP
\taxp\ derived by adding the \xmm\ flux measurements to the 
\xte\ results of \cite{ibr04}. For comparison, the \xmm\ fluxes were
extracted assuming a simple power-law spectral model fitted in the
$2-10$~keV energy band. The combined data points are well fitted by an
exponential decay model with overall time-constant of $233.5$~d, somewhat
shorter but consistent with the initial \xte\ trend, given the
extended monitoring interval. To allow for a systematic offset in the 
\xte\ flux measurements\footnote{\cite{ibr04} assumed a nominal
$2-10$~keV conversion factor of 2.27 counts~s$^{-1}$ per CPU~ $= 2.4
\times 10^{-11}$~erg~s$^{-1}$~cm$^{-2}$.}, these data were rescaled by a factor of 1.42 to 
match the \xmm\ results.  The implied flux range for
the initial outburst is $F(2-10 \ \rm{keV}) = (11 - 8) \times
10^{-11}$~erg~s$^{-1}$~cm$^{-2}$, over the interval during which the
target was inaccessible to \xte. Clearly the flux in this band has all
but returned to its pre-outburst value.

\begin{table*}
\small
\label{tab:table}       
\caption{\bf \xmm\ Spectral Results for \taxp}
\footnotesize
\begin{tabular}{lccccccc}
\hline\noalign{\smallskip}
{Parameter}           & {2003 Sep 8} & {2003 Oct 12} & {2004 Mar 11}& {2004 Sep 18}& {2005 Mar 18}& {2005 Sep 20}& {2006 Mar 18}\\[3pt]
\tableheadseprule\noalign{\smallskip}
Expo (ks)$^a$         &     11.5/8.1         &     6.9/6.2          &     17.0/15.8       &     26.5/24.4       &     39.8/37.2        &     39.5/37.8        &     41.7/38.8\\
$N_{\rm H}$~(cm$^{-2}$)$^b$         & $0.65 \pm 0.04$      & $0.65 \pm 0.04$      & $0.65 \pm 0.04$     & $0.65$~(fixed)      & $0.65$~(fixed)       & $0.65$~(fixed)       & $0.65$~(fixed)\\
$kT_1$ (keV)	      & $0.25 \pm 0.02$      & $0.29 \pm 0.04$      & $0.27 \pm 0.02$     & $0.25 \pm 0.01$     & $0.22 \pm 0.01$      & $0.20 \pm 0.01$      & $0.19 \pm 0.01$\\
$kT_2$ (keV)          & $0.68 \pm 0.02$      & $0.71 \pm 0.03$      & $0.70 \pm 0.01$     & $0.67 \pm 0.01$     & $0.60 \pm 0.01$      & $0.52 \pm 0.01$      & $0.46 \pm 0.02$\\
$A_1$ (cm$^2$)	      & $5.6\times10^{12}$   & $2.9\times10^{12}$   & $3.3\times10^{12}$  & $4.0\times10^{12}$  & $4.9\times10^{12}$   & $6.6\times10^{12}$   & $7.2\times10^{12}$ \\
$A_2$ (cm$^2$)        & $2.8\times10^{11}$   & $2.2\times10^{11}$   & $1.3\times10^{11}$  & $8.7\times10^{10}$  & $6.0\times10^{10}$   & $3.7\times10^{10}$   & $3.6\times10^{10}$ \\
BB1 Flux$^c$          & $4.2\times10^{-12}$  & $5.4\times10^{-12}$  & $3.5\times10^{-12}$ & $2.6\times10^{-12}$ & $1.6\times10^{-12}$  & $1.0\times10^{-12}$  & $7.5\times10^{-13}$\\
BB2 Flux$^c$	      & $3.5\times10^{-11}$  & $3.0\times10^{-11}$  & $1.8\times10^{-11}$ & $1.0\times10^{-11}$ & $4.0\times10^{-12}$  & $1.3\times10^{-12}$  & $6.8\times10^{-13}$  \\
Total Flux$^c$	      & $3.93\times10^{-11}$ & $3.84\times10^{-11}$ & $2.13\times10^{-11}$& $1.29\times10^{-11}$& $5.67\times10^{-12}$ & $2.35\times10^{-12}$ & $1.44\times10^{-12}$ \\
$L_{\rm BB1}$(bol)$^d$& $2.4\times10^{34}$   & $2.3\times10^{34}$   & $1.7\times10^{34}$  & $1.6\times10^{34}$  & $1.2\times10^{34}$   & $1.0\times10^{34}$   & $8.6\times10^{33}$ \\
$L_{\rm BB2}$(bol)$^d$ & $6.3\times10^{34}$  & $5.7\times10^{34}$  & $3.1\times10^{34}$  & $1.8\times10^{34}$  & $7.9\times10^{33}$  & $2.8\times10^{33}$  & $1.7\times10^{33}$ \\
$\chi^2_{\nu}$(dof)   & 1.1(187)             & 1.1(84)              & 1.1(194)            & 1.2(188)            & 1.6(152)             & 1.6(80)              & 1.6(117)\\[0.5em]
\noalign{\smallskip}\hline
\end{tabular}\\
{ {N}{\tiny OTE} --  \footnotesize Uncertainties in spectral parameters are 90\% confidence for two interesting parameters.}\\
$^a${\footnotesize EPIC-pn exposure/livetime in units of ks.}\\
$^b${\footnotesize Interstellar hydrogen absorbing column density in units of cm$^{-2}$.}\\
$^c${\footnotesize Absorbed 0.5--10 keV flux in units of erg cm$^{-2}$ s$^{-1}$.}\\
$^d${\footnotesize Unabsorbed bolometric luminosity in units of erg s$^{-1}$ assuming a distance of $d = 3.3$~kpc.}\\
\end{table*}

\begin{figure*}[t]
\centerline{
\includegraphics[angle=270,width=0.46\textwidth,clip=true]{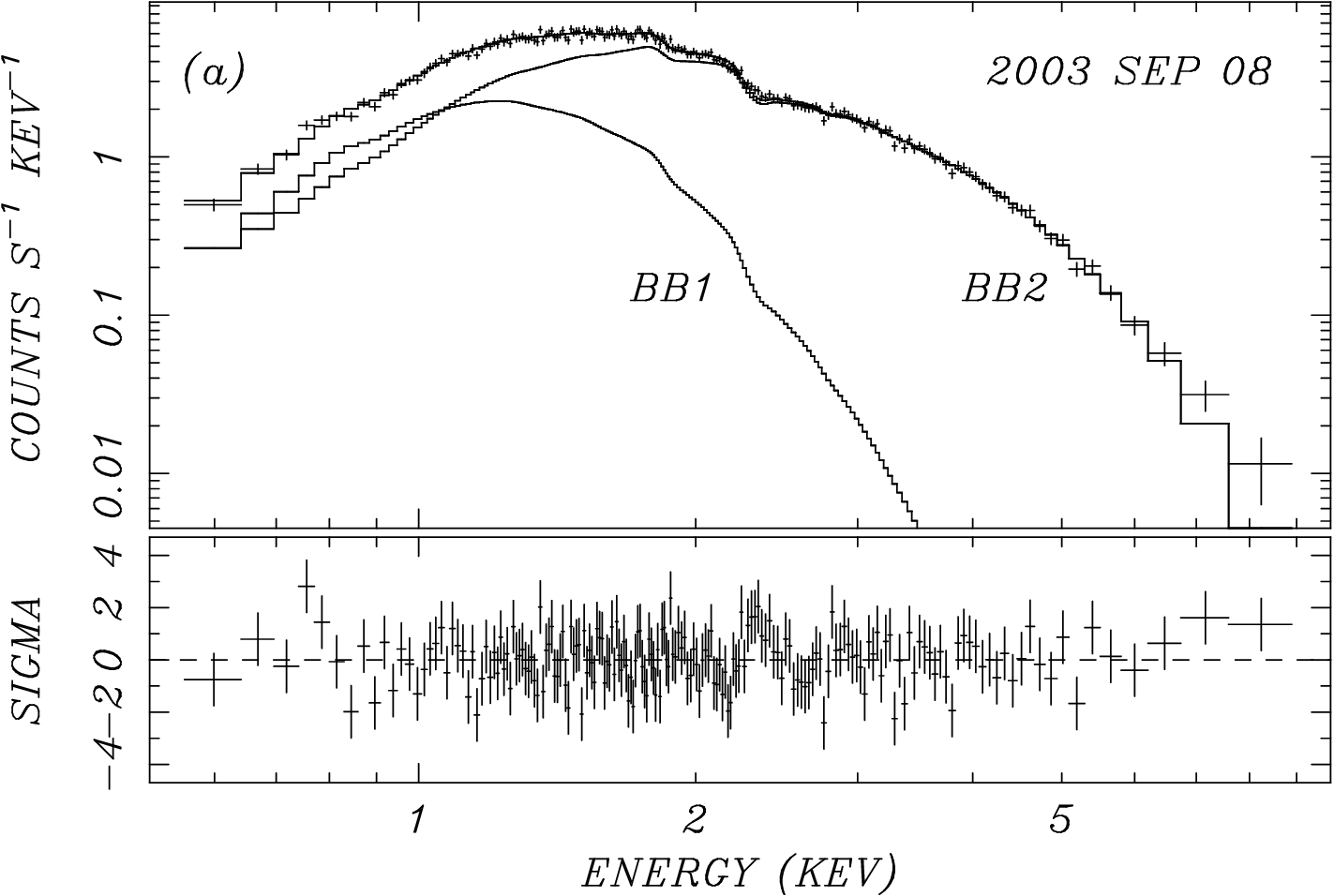}
\hfil
\includegraphics[angle=270,width=0.46\textwidth,clip=true]{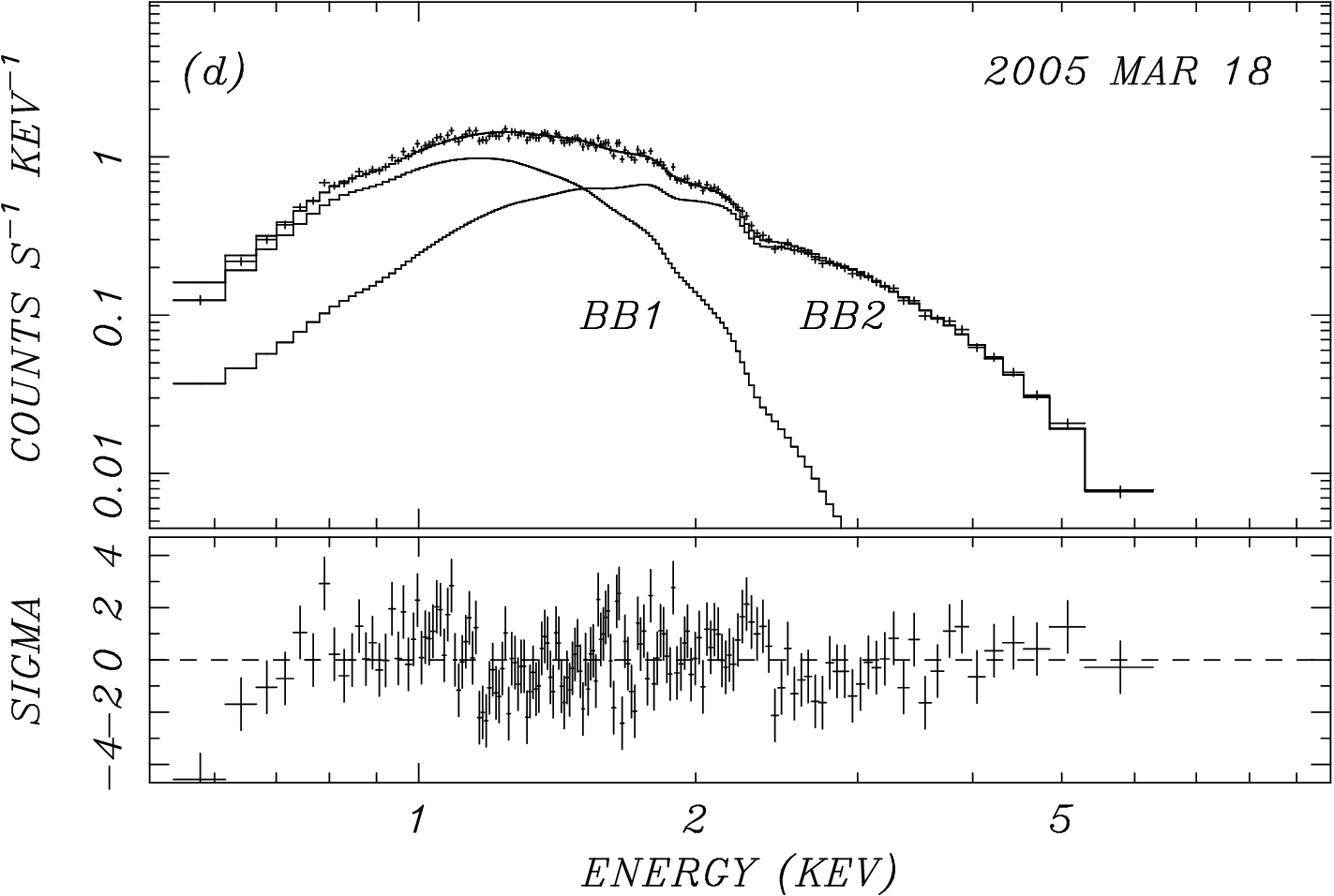}
}
\vspace{0.25cm}
\centerline{
\includegraphics[angle=270,width=0.46\textwidth,clip=true]{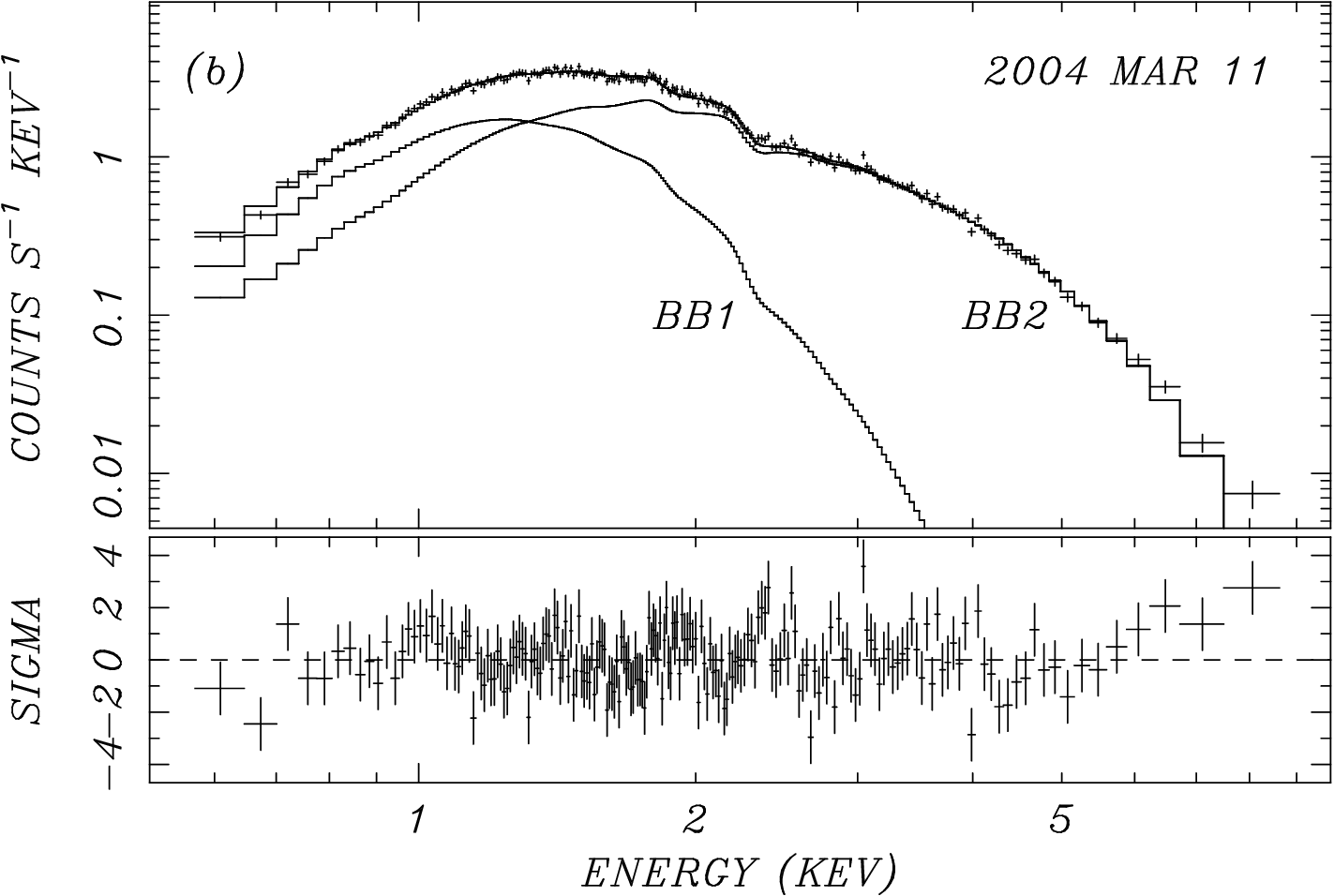}
\hfil
\includegraphics[angle=270,width=0.46\textwidth,clip=true]{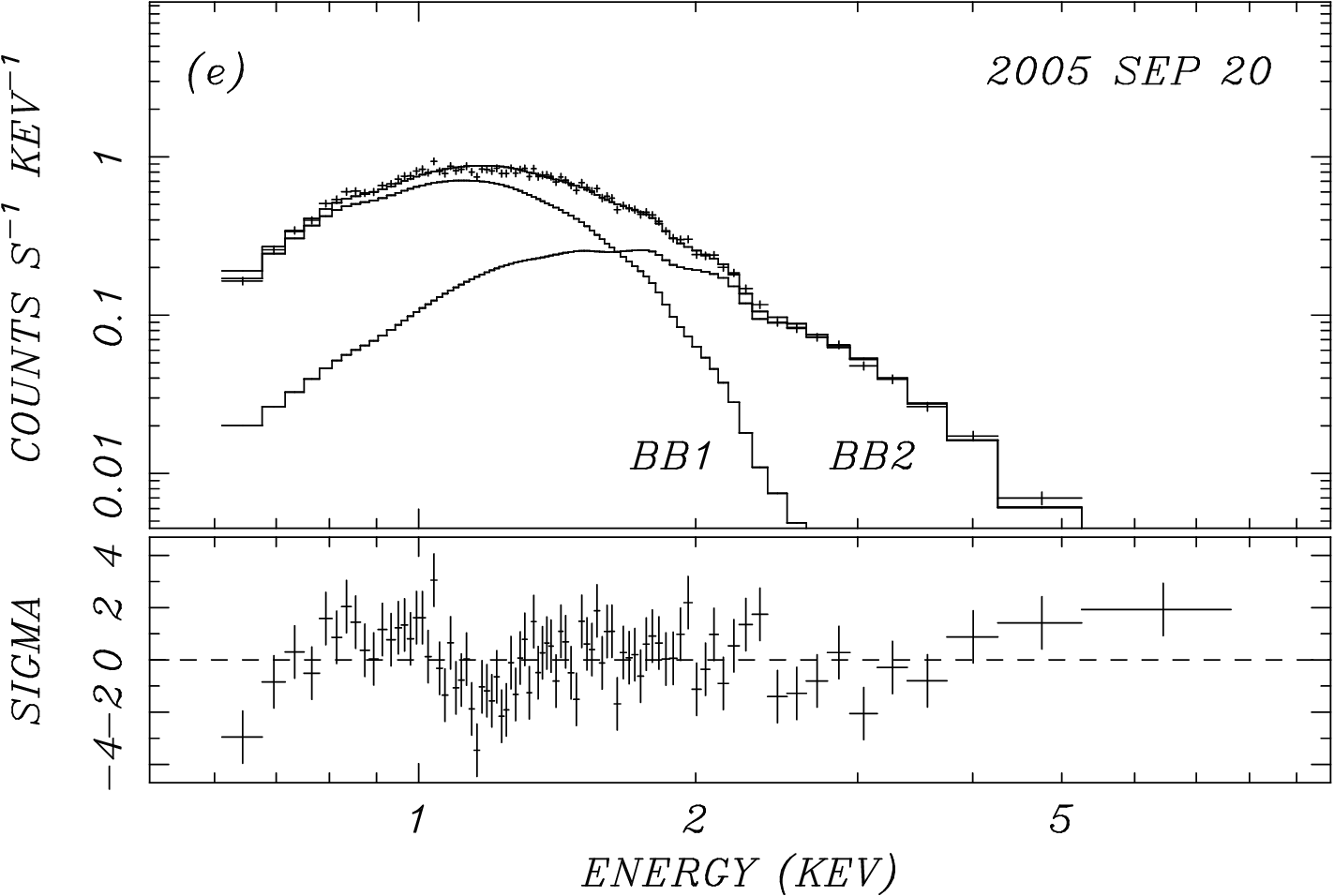}
}
\vspace{0.25cm}
\centerline{
\includegraphics[angle=270,width=0.46\textwidth,clip=true]{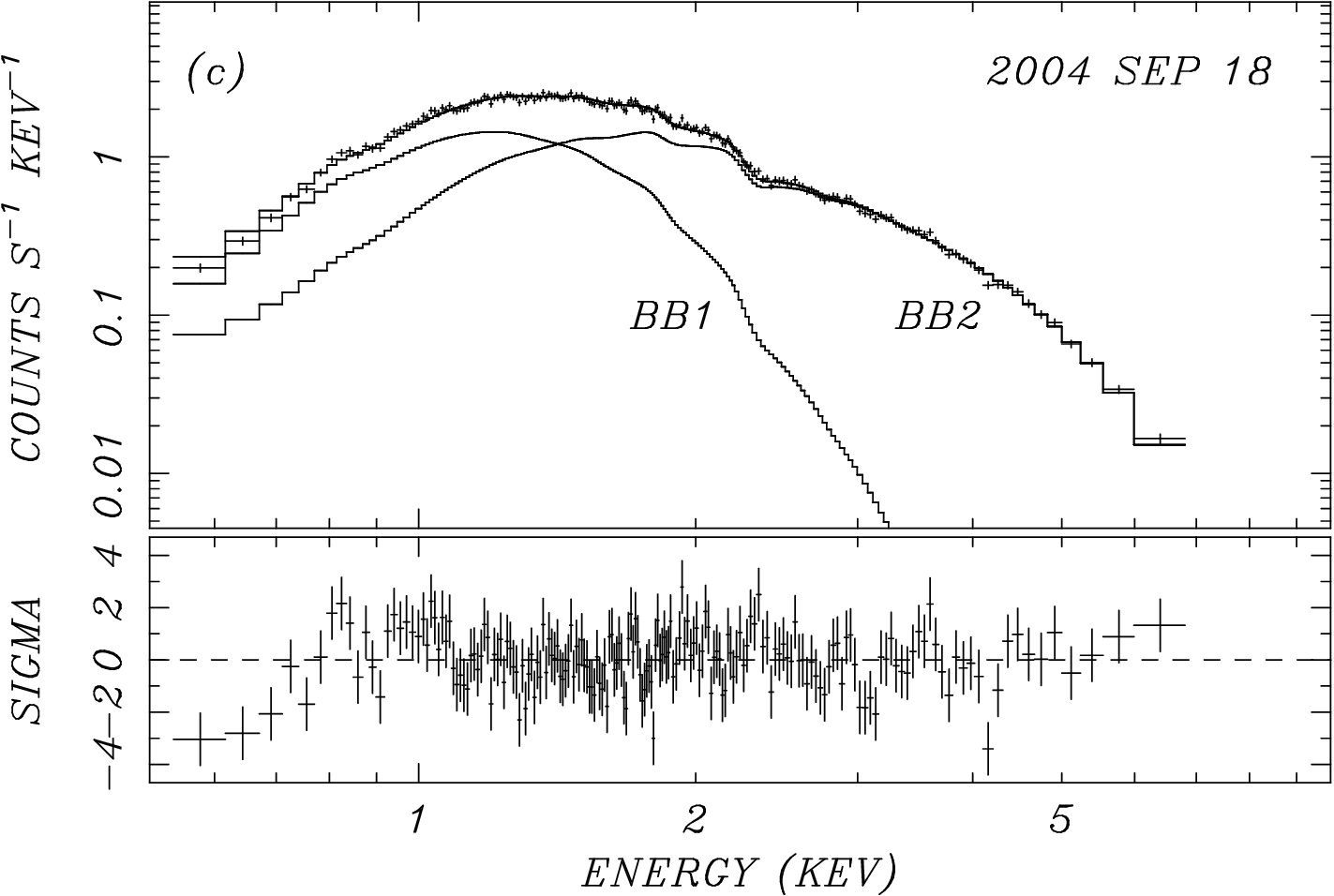}
\hfil
\includegraphics[angle=270,width=0.46\textwidth,clip=true]{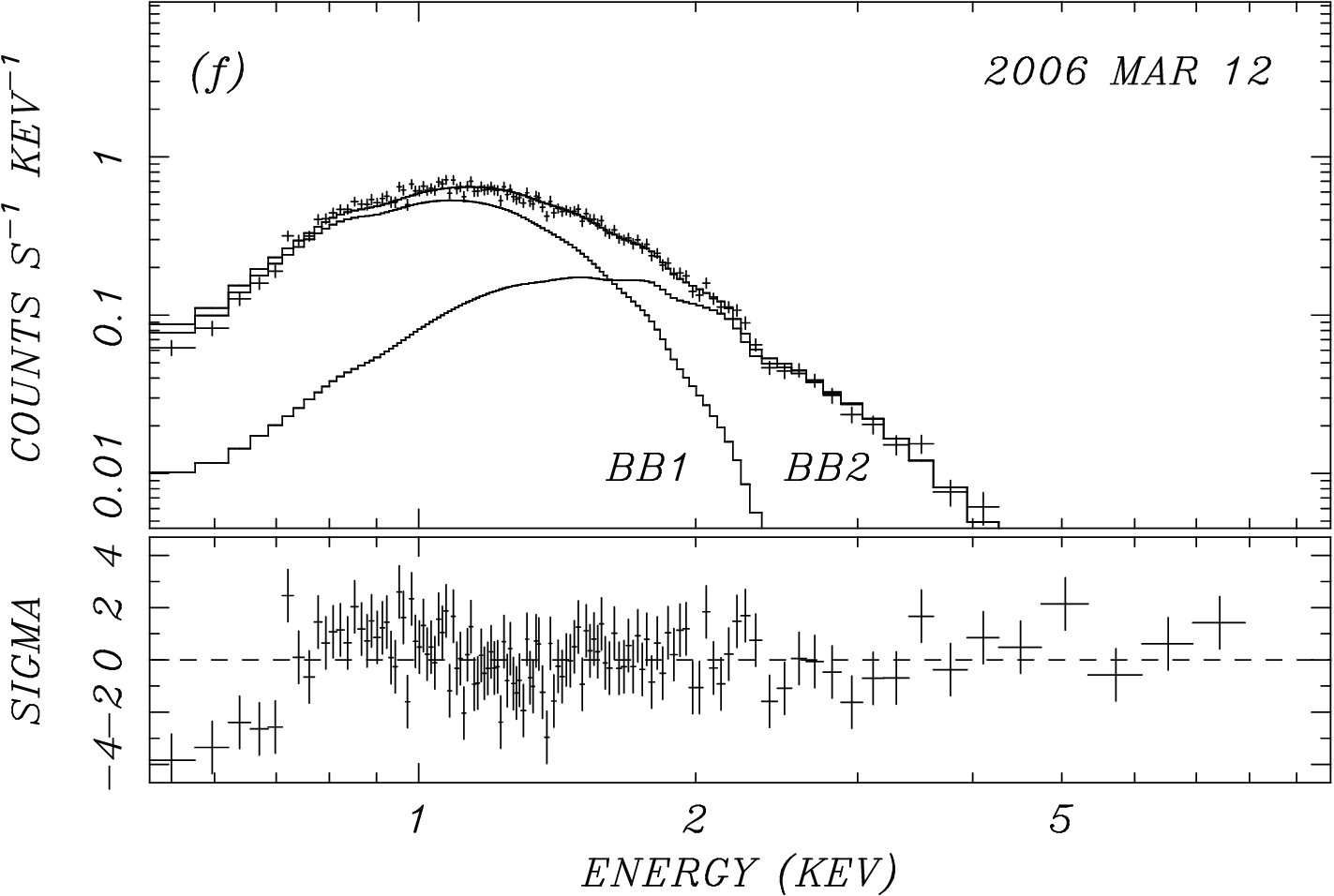}
}
\caption{\xmm\ EPIC~pn spectra of \taxp\ from the earliest to the 
latest epochs, in six months intervals (the 2003 Oct spectrum is
excluded, for clarity). These spectra are shown with the best-fit
two-temperature blackbody model specified in Table~1.  Although the
temperatures of the blackbody components have not changed greatly
between epochs, the flux of the hot component (BB2) has decayed
rapidly relative to that of the cooler one (BB1). Also shown are the
residuals from the best-fit models. The nature of the deviations to
the model below $0.7$~keV has yet to be determined (see text).
\label{fig:spectra}}
\end{figure*}

\begin{figure}[t]
\centerline{
\includegraphics[angle=0,width=0.45\textwidth,clip=true]{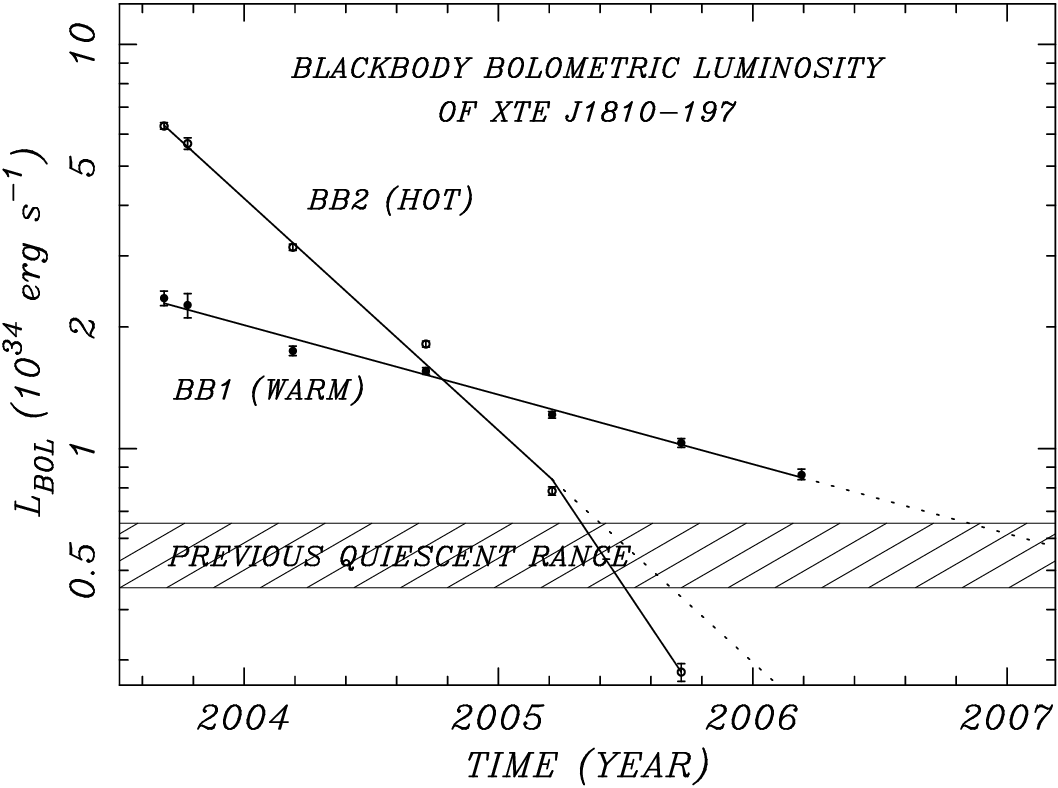}
}
\caption{A semi-log plot of the bolometric luminosity  of \taxp\ as a
function of time ({\it crosses}) for each temperature component of the
double blackbody model given in Table~1. The light-curves
are fit assuming an exponential decays model ({\it solid lines}); the
corresponding e-folding times is $\tau_1 = 280$~d and $\tau_2 = 870$~d
for the warm and hot temperature components, respectively. The recent
data for the hot component now deviates from this model.  The fitted
quantities have been extrapolated to the ($1\sigma$) quiescent range
measured in Paper~I ({\it cross-hatched area}).
\label{fig:luminosity}}
\end{figure}

Spectra of AXPs are nominally modeled assuming a two-component
power-law plus blackbody model. The spectral and temporal trends for
\taxp\ allow us to strongly reject this model (see
Paper~II for a comprehensive discussion). Instead, we find that a
two-temperature blackbody model gives equally acceptable spectral fits
but is better motivated physically (see \S5). A summary of spectral
results is presented in Table~1 and the spectra, fitted with the
double blackbody model, are displayed in Figure~\ref{fig:spectra} (the
2003 Oct observation is excluded here for clarity). Although the hot
blackbody component (BB2; Fig.~\ref{fig:luminosity}) initially
dominated the emission at low energies, it fades more rapidly than the
cooler ``warm'' emission component (BB1; Fig.~\ref{fig:luminosity}).
Evidently the spectral fits after 2004 March deviate significantly at
the lower end of the \xmm\ energy band, below 0.7~keV. It is not yet
clear if this is an instrumental artifact, however, taken at face
value there seems to be an absorption feature at $\approx 350$~eV with
a Gaussian width of $\sigma \approx 160$~eV. The mean equivalent width
of this feature is $~600$~eV, but its strength is somewhat time
dependent. Perhaps as the hotter component contributes less, this
feature becomes more evident, suggesting an association exclusively
with the cooler component.  Further research is needed to quantify
this feature and consider its reality, or physical implications.

%

The spectrum of \taxp\ can be thought of as the combined flux from two
concentric emitters, ``hot spots'', whose temperature and size are
evolving at different rates, effectively changing the overall shape of
the spectrum with time (Paper II). With a set of spectral measurements
spanning two and a half years, a clear trend has emerged.  The
bolometric luminosities of the two components are shown in
Figure~\ref{fig:luminosity}. This reveals a cooler component
decreasing exponentially in time with $\tau_1 = 870$~d, while the
hotter temperature flux declines with $\tau_2 = 280$~d.  As might be
expected, the shorter time constant is very close to the one described
by the long-term flux above $2$~keV (Figure~\ref{fig:flux}), where the
hotter blackbody component dominates. However, the latest two data
points show that this component has now fallen faster than the
original exponential. Although the {\it bolometric luminosities} of the two
components are quite different at the latest epoch, their measured
{\it fluxes} are now nearly equal in the $0.5-8$~keV range. We also
note that, while it is possible to fit an alternative power-law decay
model to the hot blackbody flux, such a model would require a decay
index that steepens with time.

Based on the decay rates and approximate outburst time we estimate an
initial bolometric luminosity for the two spectral components of
$L_1 \approx 3 \times 10^{34} \,d^2_{3.3}$~erg~s$^{-1}$ and $L_2
\approx 2 \times 10^{35} \,d^2_{3.3}$~erg~s$^{-1}$, respectively. The
total luminosity at the peak of the outburst is comparable to that of
a persistent AXP, suggesting that this TAXP is anomalously faint.  We
can now compute the implied fluences of $f_1
\approx 2 \times 10^{42} \,d^2_{3.3}$~erg and $f_2 \approx  4 \times 
10^{42} \,d^2_{3.3}$~erg, for the two components, respectively.  The
energy of the outburst event is many orders-of-magnitude lower than
that for a typical SGR flare.

\begin{figure}[t]
\centerline{
\includegraphics[angle=270.0,width=0.45\textwidth,clip=true]{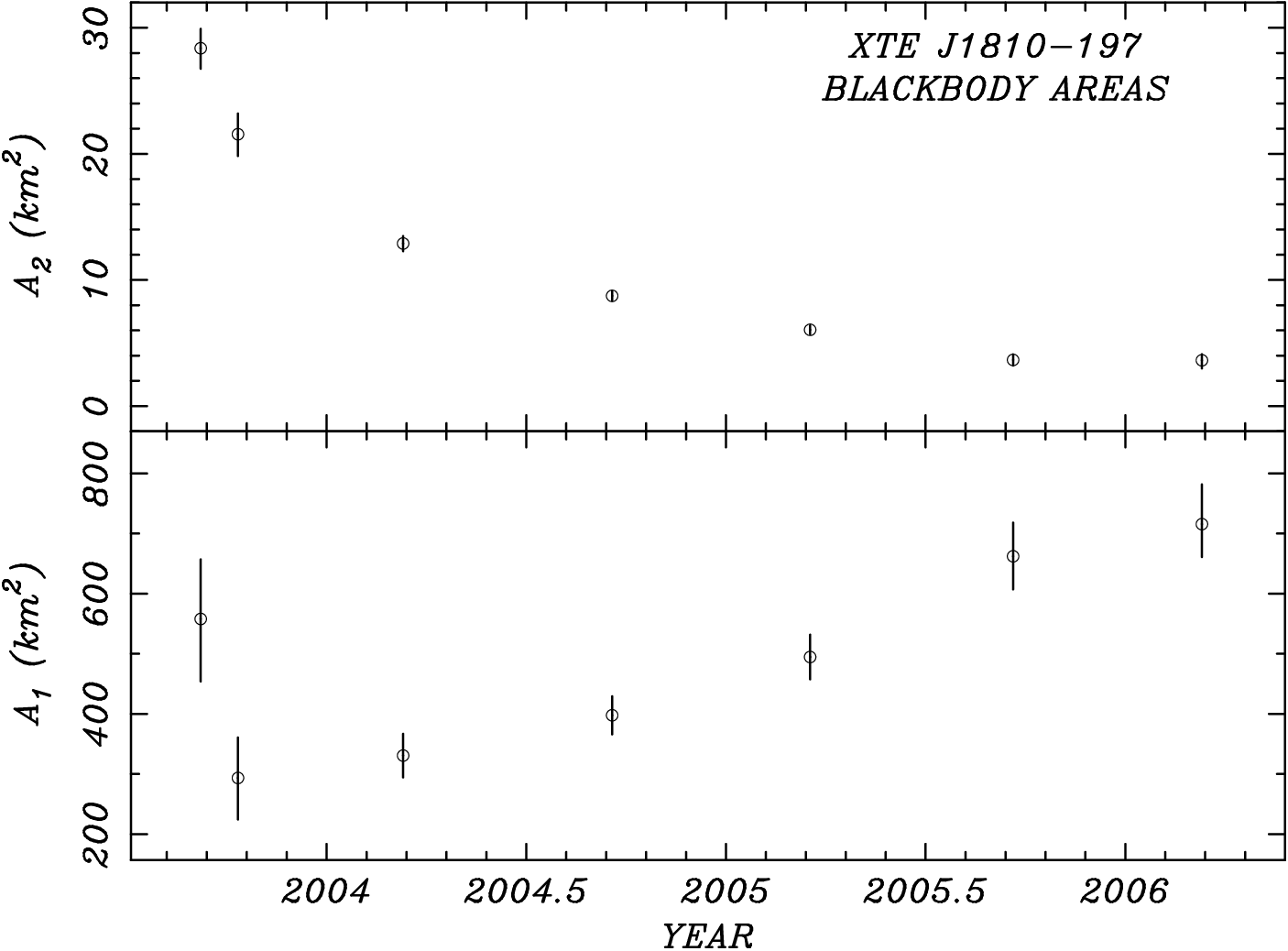}
}
\vskip 0.15in
\centerline{
\includegraphics[angle=270.0,width=0.45\textwidth,clip=true]{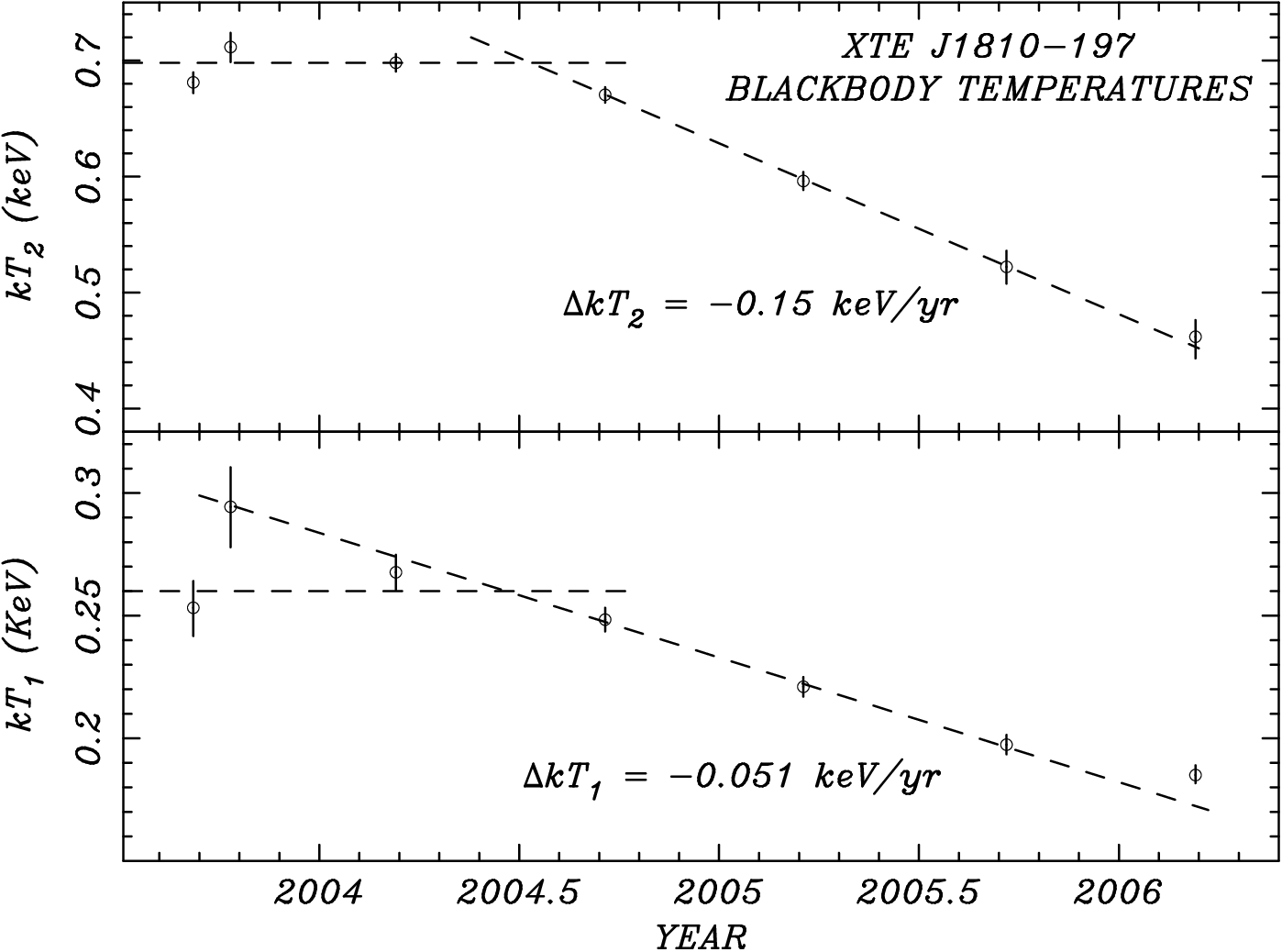}
}
\caption{
The time history of the best-fit \xmm\ component temperatures and
surface emitting areas for \taxp\ based on the double blackbody
spectral model (Table~1).  The evolving areas ({\it Top})
indicates that the warm component is expanding to cover the NS
surface, while the hot component is shrinking rapidly. The
corresponding temperatures ({\it Bottom}) are declining at a steady
rate of $22\%$ per year since mid 2004.  Prior to that time the
temperature of the hotter component was essentially constant. This
trend for the warm component is somewhat ambiguous.
\label{fig:params}
}
\end{figure}

The temperatures and inferred blackbody emitting areas have also
evolved measurably with time. Figure~\ref{fig:params} displays the
time histories of these parameters for each spectral component,
obtained from the model fits. Based on the three latest data points,
the temperatures now show a definite cooling trend for each spectral
component, as suggested by the broken line in Figure~\ref{fig:params}.
Prior to mid 2004, the temperatures likely remained nearly constant,
after which they both fell at a rate of $\approx 22\%$ per year
($\Delta kT_1 = -0.051$~keV~yr$^{-1}$; $\Delta kT_2 =
-0.15$~keV~yr$^{-1}$).  The size of the effective hot spots (derived
from the blackbody emission areas) also followed distinct
evolutions. The hot component has been shrinking exponentially since
the initial \xmm\ observation, while the warm component has steadily
increased in size. A notable exception is the initial data point which
deviates from the trend for both the areas and temperatures. This may
be associated with a glitch or rotation instability, as suggested by the
pulse timing results around this epoch (\S3).

The area of the warm component ($A_1$; Table~1) may be reaching a
maximum, corresponding to the whole NS surface. This would then provide a
lower-limit on the NS radius of $\simgt 7.5 \,d_{3.3}$~km, ignoring
relativistic effects (redshift and light bending). In contrast, the
area of the hotter component continues to shrink and its contribution
to the flux is severely diminished.

As shown in Figure~\ref{fig:luminosity}, the X-ray luminosity of TAXP
\taxp\ has nearly returned to its historic quiescent level.  This is
also true of the temperature and inferred blackbody area.  The
pre-burst \rosat\ spectrum of 1992 March 7 is reasonably well fitted
with a single blackbody of $kT = 0.18\pm 0.02$~keV covering
$5.2\times10^{12}\,d^2_{3.3}$~cm$^2$ and $L_{BB}({\rm bol}) = 5.6
\times 10^{33}\,d^2_{3.3}$~erg~s$^{-1}$ (Paper I). The latest \xmm\
spectrum that overlaps the \rosat\ band is again dominated by the
cooler blackbody component. Furthermore, the current pulsed fraction
is consistent with the quiescent upper-limits. Thus the double
blackbody model is sufficient to describe the observed properties, and
may not require a third ``quiescent'' component to explain the
pre-outburst observations. \xmm\ data from the next observing window
(2006 September) is likely to be consistent with the \rosat\ result in
the soft energy band, effectively defining a return to the quiescence state.

\section{Spin-down Evolution of \taxp}

\begin{figure}
\centerline{
\includegraphics[angle=270.0,width=0.45\textwidth,clip=true]{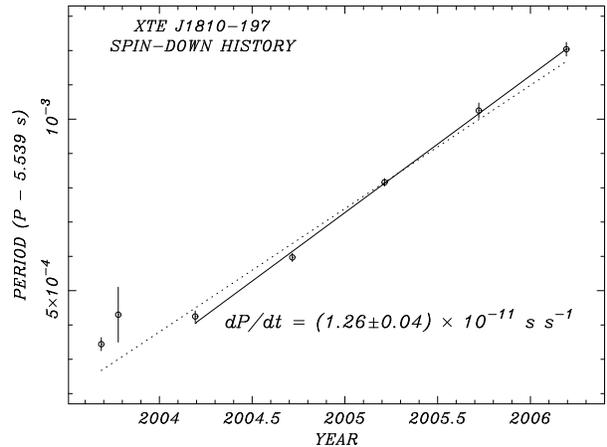}
}
\caption{The spin-down history of \taxp\ pulse period as measured using the 
\xmm\ data of Table~1. The initially erratic spin-down appears 
to be settling down -- the last 5 data points are well-modeled by a
linear spin-down model ({\it solid line}); in contrast, a poor fit is
obtained using all points ({\it dotted line}).
\label{fig:spindown}
}
\end{figure}

The barycentered pulse period of \taxp\ measured at each \xmm\
observing epoch is shown in Figure~\ref{fig:spindown}. These data
points were derived following the method outlined in Paper~I.  Prior
to epoch 2004, the spin-down rate derived from the \xte\ and initial
\xmm\ measurements were highly erratic, ranging from $\dot P = (0.8-2.2)
\times 10^{-11}$~s~s$^{-1}$ (\citealt{ibr04}; Paper~III). This temporal 
behavior is likely associated with the outburst event.  The latest
period measurements suggest that the spin-down rate has settled down
somewhat. A linear model fit to the last 5 period measurements yields
a period derivative of $\dot P = 1.26\pm0.04 \times
10^{-11}$~s~s$^{-1}$ with a $\chi^2_\nu = 1.2$ for 3 DoF,
corresponding to a null hypothesis probability of 0.27. Compared to
the radio $\dot P = 1.016\pm0.001 \times 10^{-11}$~s~s$^{-1}$ of 17
March - 7 May \citep{cam06}, there is evidently still timing noise in
the pulsar's spin-down. The revised X-ray rate implies a nominal
characteristic age $\tau_{\rm c} = 7.0$~kyr, surface magnetic field
$B_{\rm s} = 2.7\times 10^{14}$~G, and spin-down power $\dot E = 2.9
\times 10^{33}$ erg~s$^{-1}$, values typical for a magnetar.
%

\begin{sidewaysfigure*}
\centerline{
\includegraphics[angle=270.0,width=0.9\textwidth,clip=true]{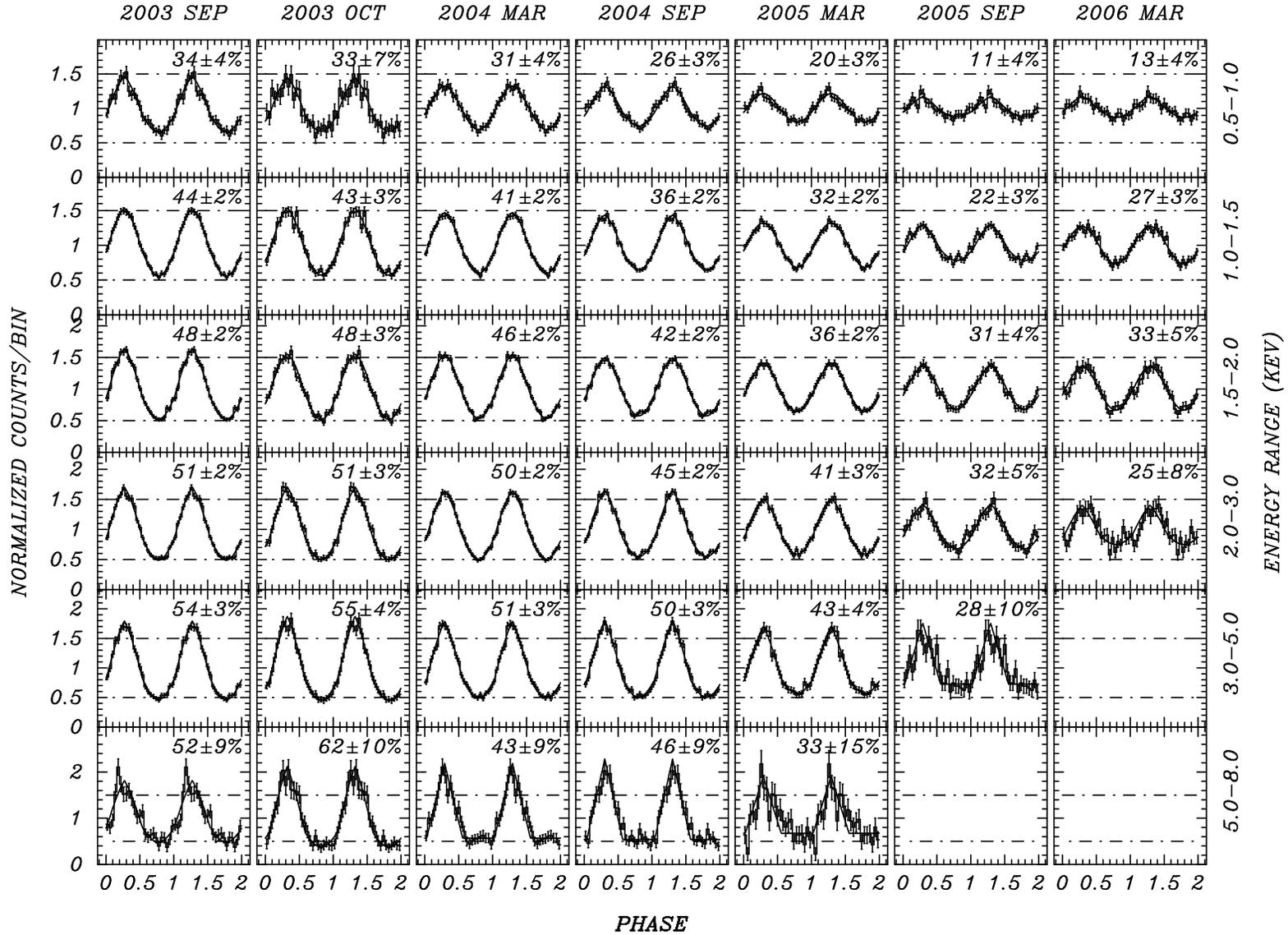}
}
\caption{Energy-dependent pulse profiles of \taxp\ obtained with the
\xmm\ EPIC~pn detector for all seven epochs of Table~1. The
background for each profile has been subtracted and phase zero is
aligned for each epoch, for comparison.
The phase of the peak is seen to be energy independent.
The pulsed fraction at low X-ray energies has decreased with time,
while remaining essentially unchanged at high energy. Also shown
is the best fit to the two-component model for the pulse profile ({\it
solid line\/}) described in the text (see \S 2.1).
\label{fig:profiles}
}
\end{sidewaysfigure*}

The energy-dependent modulation of the pulsar has evolved noticeably
over time with the declining flux. Figure~\ref{fig:profiles} displays
the pulse profiles at each \xmm\ epoch, folded at the best determined
periods. The data has been broken up into six energy bands, with the
corresponding background in each band subtracted.  The pulsed
fractions, defined as the pulsed emission divided by the total flux,
are indicated on the corner of each panel of the figure.  These
profiles are aligned so that phase zero is the same at each epoch, for
ease of comparison. In any case, within an observation, the phase zero
does not change with energy. This suggests a geometric interpretation
for the modulation, such as from localized emission on the rotating NS
star surface. The pulsed fraction has generally decreased with time,
most notably at X-ray energies below $E<2$~keV; at higher energies the
trend is less clear due to the large uncertainties derived for those
pulsed fractions. Furthermore, at all epochs, the pulsed fraction
clearly increases with energy, an equally interesting result discussed
below.

The evolving spectral shape and pulse profiles of TAXP \taxp\ provide
an important and unique (so far) diagnostic of the emission geometry,
and ultimately, the emission mechanism(s) of NSs.  Because the
spectrum is well modeled by two blackbody components, and the phase
alignment of the pulse profiles are energy independent, it is most
natural to consider emission from two concentric regions on the NS
surface. The hot component is associated with a smaller hot-spot,
while the blackbody model predicts a larger annulus for the warm
component. The complete collection of observed pulsed fractions is
consistent with this model.  The smaller, hotter spot always dominates
the spectrum above $2$~keV and offers a natural explanation for the
higher modulation at these energies. However, at lower energies, this
contribution gradually fades relative to cooler emission (which is
less modulated), contributing less and less to the pulsed emission
over time.

We note in passing that a fitted power-law spectral component would have to
dominate the soft ($< 2$~keV) X-rays at {\it all\/} epochs, while
varying in its contribution to the hard ($> 2$~keV) X-rays. Such a
power law would drive an evolution of the pulse shapes that is
opposite of what is observed.  Thus, we find that the detailed
evolution of the X-ray emission from \taxp\ further supports the
assumption of a purely thermal spectral model, and leaves no evidence
of a steep power law as is commonly fitted to individual observations
of AXPs.

Since the first observation of \taxp\ it was apparent that the
broad-band pulse shape is not a simple sine function; the pulse peak
is relatively sharp with a broader inter-pulse trough (Paper~I). This
effect is more pronounced at higher energy, where the profile is
nearly triangular in shape. This suggests that the pulse profiles can
be decomposed into a triangle function and a sinusoidal function, a
model that was explored for the earlier data sets and detailed in
Paper~II \& III. This model continues to be appropriate for the new data
and is used to extract unbiased pulsed fraction measurements for the
profiles shown in Figure~\ref{fig:profiles}.

Given the success in modeling the pulse profiles with these functions
it is natural to associate the two pulsed components uniquely with the
two spectral components, i.e., the triangle shape for the hotter
blackbody component and the sinusoidal shape for the cooler one.
However, we find that we can not model all the profiles in a
consistent manner with just a simple superposition of these temporal
components, based on the implied flux ratio in each energy band
\citep{got05}. Instead we conclude that either the two
spectral components contain an admixture of the two shapes or there
is a third, unmodeled spectral component present. This is a direction of
active research, to consider the correct characterization of the phase
dependent spectral contribution.

\section{Radio observations of \taxp}

\begin{figure}[t]
\centerline{
\includegraphics[angle=270.0,width=0.45\textwidth,clip=true]{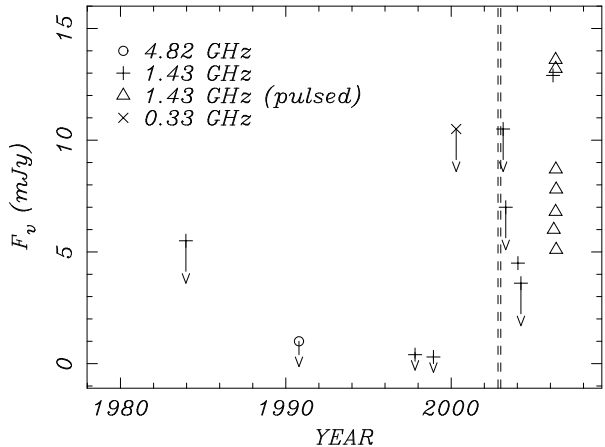}
}
\caption{
A set of archival and dedicated radio observations of \taxp\ measured
at three frequencies.  Prior to outburst (denoted by vertical lines)
no radio emission is found corresponding to the plotted
upper-limits. Following the X-ray outburst, large variations in flux
are inferred from several measurements and upper-limits. Pulsed radio
emission is also recorded at the X-ray period in the post-outburst era
following the 2006 discovery. A search of archival data prior to the
outburst places an upper-limit of $\sim 0.4$~mJy at 1.4~GHz. For
clarity, multi-frequency measurements coinciding with the 1.4~GHz
pulsed emission data points are not shown (from
\citealt{hal05b,cam06}).
\label{fig:radio}
}
\end{figure}

Considering that the absence of radio emission is a defining
characteristic of AXPs, the serendipitous radio detection from
\taxp\ came as a great surprise. A chance search of VLA data taken
about a year after outburst reveals an unresolved point source with a
flux of $4.5$~mJy at 1.43~GHz, located at the precise coordinates of the
TAXP \citep{hal05b}. Other archival VLA data obtained at various
frequencies provide upper-limits before and after outburst. A
follow-up VLA observation in 2006 March yielded a flux of $12.9$~mJy
at 1.43~GHz.  Together, these results indicate highly erratic radio
emission.  A second surprise came with the detection around that time
of pulsed radio emission at the X-ray period \citep{cam06}. This
search, at $\nu = 1.4$~GHz using the Parkes radio observatory,
revealed a narrow, bright pulse with a high degree of linearly
polarization \citep{cam06}. A series of multi-frequency measurements
in 2006 April-May confirms the erratic flux behavior and provides an
unusual spectral slope of $\alpha \simgt -0.5$, where $S_\nu
\propto \nu^{\alpha}$ (cf. $\alpha \sim -1.6$ for a typical radio pulsar). 
A revised distance to the pulsar of $3.3$~kpc is inferred from the pulsar's dispersion.
A search for pulsations in archival data acquired in 1997/1998
produced a null result and argues against significant emission prior
to the outburst event. A summary of these radio observations is shown
in Figure~\ref{fig:radio}.

\section{Emission Geometry: Models \& Theory}

Although transient AXPs such as \taxp\ are relatively rare, their
short active duty cycle suggests the existence of a larger population
of unrecognized young NSs. \taxp\ provides a unique window into this
population, with prior measurements in the quiescent state and
detailed observations during its active, pulsed phase. Ultimately, we
hope to use this TAXP to give insight into the emission mechanisms
of magnetars, in general. With the discovery of pulsed radio flux,
this may carry over to interpreting emission mechanism of radio
pulsars, as well.  Below we now summarize our initial attempts to
interpret the \taxp\ results in the context of a physical model.
First we outline the basic arguments for the two-temperature blackbody
spectral model as a plausible alternative to the nominal power-law
plus blackbody model.

While both models give reasonable fits to the spectrum, it's important
to note that the power-law component is used to model the {\it low
energy residuals and not any high-energy tail}. Without a low energy
cut-off of this component, it is not possible to connect it with the
observed IR emission without an energy catastrophe. As detailed in
Paper~II, invoking synchrotron self-absorption is excluded by
radius/magnetic-field inconsistently.  In contrast, the extrapolated
spectrum of the warm blackbody component does not exceed the
measured IR flux. Furthermore, currently there is no acceptable
physical model to anchor a power-law component. As discussed earlier,
the two components of the double blackbody model can be associated
with a pair of hot spots on the surface of the star. The warm
temperature component, covering a large fraction of the NS surface, is
consistent with the decreased pulsed fraction at lower energies. The
hotter blackbody component is consistent with a smaller emission area
and greater modulation at higher energies.  A comprehensive discussion
of this issue can be found in paper~II.

Determining the emission geometry on the NS is of great interest. The
natural conclusion of applying the double blackbody model to the time
resolved spectra and pulse profiles is that concentric hot regions
give rise to the observed modulation. We consider a model of the
phase-resolved spectrum taking into account the viewing geometry and
offset of the hot spots from the rotation axis (Perna \& Gotthelf
2006, in prep.). This is based on the NS emission model given in
\cite{per01} 
that includes general relativistic effects (redshift and light
bending). Our goal is to match spectrum and energy dependent pulse
shape in order to determine viewing geometry, distance, and NS
size. Preliminary work is able to reproduce the pulsed fraction to a
reasonable degree but the pulse shape remains elusive.

A framework for a theoretical interpretation of the emission from
\taxp\ is suggested by the magnetar coronal model of
\cite{bel06}. According to this model, the large outburst was generated
by a starquake that resulted in a transition to an active coronal
state that caused energy to be stored in the twisted B-field of the
coronal loop. Particles (mostly $e^+e^-$) are accelerated within this
loop and impact the NS surface with GeV energy. This heats up the loop
footprint resulting in the observed hot spot emission. The decay
timescale of the coronal loop, and thus the hotter component of the
double blackbody luminosity, is of order of a few years. The decay
rate is determined by ohmic dissipation of current in the excited
loop. A cooler component likely arises from deep crustal heating
associated with the initial outburst and possibly earlier ones. The
features of this model are in general accordance with the
observational properties and inferred model for \taxp.

\begin{acknowledgements}

We thank Fred Jansen and Norbert Schartel for providing the four \xmm\
Targets of Opportunity observations of \taxp. 

\end{acknowledgements}

\end{document}